\documentclass[reprint,superscriptaddress, amsmath,amssymb,aps,prl]{revtex4-1}

\usepackage{graphicx}
\usepackage{bm}
\usepackage{color}

\begin{document}

\title{Drying Pathways of an Evaporating Soft Matter Droplet}

\author{Guangle Du}
\affiliation{CAS Key Laboratory of Theoretical Physics, Institute of Theoretical
Physics,Chinese Academy of Sciences, Beijing 100190, China}
\affiliation{Wenzhou Institute, University of Chinese Academy of Sciences, Wenzhou, Zhejiang 325001, China}
\affiliation{School of Physical Sciences, University of Chinese Academy of Sciences, Beijing 100049, China}

\author{Fangfu Ye}
\affiliation{Wenzhou Institute, University of Chinese Academy of Sciences, Wenzhou, Zhejiang 325001, China}
\affiliation{School of Physical Sciences, University of Chinese Academy of Sciences, Beijing 100049, China}
\affiliation{Beijing National Laboratory for Condensed Matter Physics, Institute of Physics, Chinese Academy of Sciences, Beijing 100190, China}
\affiliation{Songshan Lake Materials Laboratory, Dongguan, Guangdong 523808, China}

\author{Masao Doi}
\affiliation{Center of Soft Matter Physics and its Applications, Beihang University, Beijing 100191, China}

\author{Fanlong Meng}
\email{fanlong.meng@itp.ac.cn}
\affiliation{CAS Key Laboratory of Theoretical Physics, Institute of Theoretical
Physics,Chinese Academy of Sciences, Beijing 100190, China}

\date{\today}

\begin{abstract}
Micro-droplets of soft matter solutions have different morphologies upon drying,
and can finally become wrinkled, buckled or cavitated particles.
We investigate the morphology evolution of a drying soft matter droplet in this work:
at the early stage of drying, wrinkling or cavitation instability can occur in the droplet, depending on the comparison between the critical wrinkling and cavitation pressure;
at a later stage of drying, no wrinkle will appear if cavitation happens first, while there can still be cavitation if wrinkling happens first.
A three-dimensional phase diagram in the space of elastic length, gel layer thickness and weight loss is provided for illustrating these drying pathways of a soft matter droplet, which can guide future fabrications of micro-particles with desired morphologies.
\end{abstract}

\maketitle

A soft matter droplet consisting of polymer solutions or colloidal dispersions
can exhibit different morphologies such as
buckling~\cite{tsapis2005,boulogne2013,pathak2015,tirumkudulu2018},
wrinkling~\cite{rio2006,zhang2014,vanderkooij2016,zheng2018a}, or
cavitation~\cite{arai2012,sadek2013,bansal2016,bansal2018} during the drying process.
When all solvents evaporate,
the soft matter droplets can turn out to be solid, hollow, wrinkled or buckled
particles as the final products~\cite{walton1999,walton2000,elversson2003,vehring2007,vehring2008}.
This drying process, especially spray drying, has been widely utilised to
produce micro-particles of different morphologies in industrial circumstances
such as food or pharmaceutical particle
production~\cite{vehring2008,anwar2011,giovagnoli2014,shishir2017,huang2017}, amorphous
material crystallisation~\cite{price2004,chiou2007,chiou2008}, functional encapsulated particle
manufacture~\cite{ire1998,estevinho2013,trojanowska2017}, \emph{etc.}, where
different shapes are achieved by empirically changing the drying temperature,
concentration and constitution of the soft matter
solution~\cite{iskandar2003,sen2011,lintingre2015,lintingre2016,pathak2015,pathak2016,biswas2016,raju2018}.

In the drying process of a soft matter droplet, the competition of solute deposition on the droplet surface due to drying and solute diffusion homogenising the solute concentration is captured by the dimensionless evaporating P\'eclet number. When P\'eclet number is sufficiently large, the
solution at the outmost layer of the droplet can solidify to be a gel-like
layer~\cite{Okuzono2006,vehring2007,maki2011,meng2016}, which can grow with time.
This gel layer has been believed as related with the morphology evolution and
the final configurations of drying soft matter
droplets~\cite{pauchard2003,meng2014,lintingre2016,sadek2016,bansal2018}. However, how the properties of the gel layer can determine the drying process of the soft matter droplet still remains unclear.

\begin{figure}[tb]
  \centering
  \includegraphics[width=0.9\linewidth]{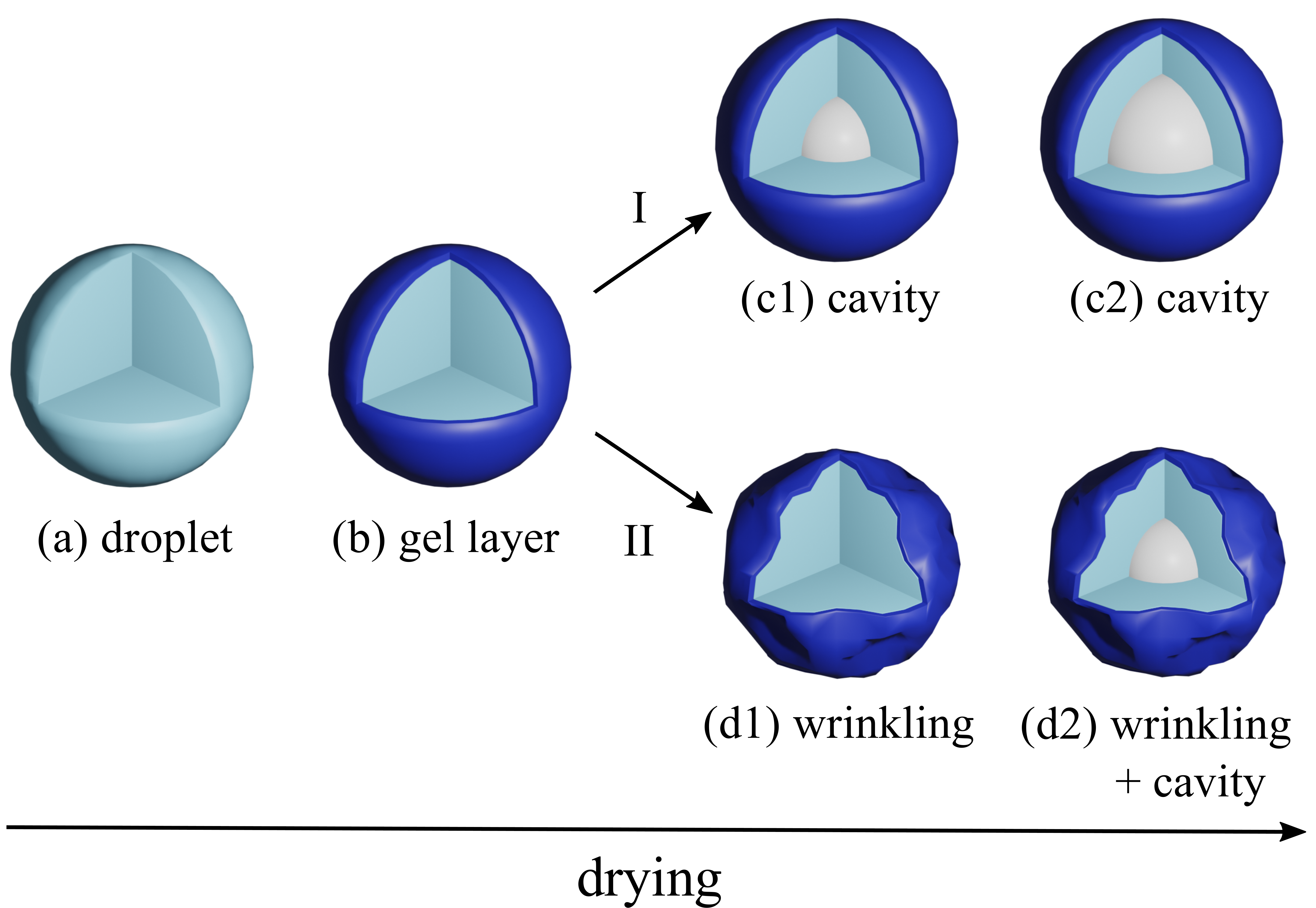}
  \caption{\label{fig:schematics_phases} Drying pathways of an evaporating soft matter droplet. }
\end{figure}

In this work,
we will study how the morphology of a drying soft matter droplet can evolve with time depending on the physical properties of the gel layer such as its elasticity and dimension, by considering the prepared state of the droplet as a spherical core-shell structured system as in spray drying [shown in Fig.~\ref{fig:schematics_phases}(a)].
The core will be simply regarded as liquid, whose amount decreases continuously due to evaporation; the shell (skin layer) will be treated with non-evolving elastic properties and thickness here for simplicity, as the evolution of the skin layer, such as mass growing from further diffusion and deposition of the solute, would not change the results qualitatively. By taking these simplifications, we can construct the analytical energy form of the system, and then discuss the morphology evolution of the drying soft matter droplet.

Compared to solvent evaporation,
the mechanically equilibrating process of the gel layer is a fast process;
in other words, we can use the mass loss of the droplet due to evaporation, $\Delta W = W_0 -W$, as the controllable variable of the system,
where $W_0$ and $W$ denote the initial and current mass of the droplet, respectively.
Then we can discuss the morphology evolution of the droplet by optimising the total free energy of the system under given $\Delta W$.
By taking the initial radius of the droplet as
$R_0$, and the radius of the cavity as $R_c$ [reduced cavity volume $v_c \equiv
{(R_c/ R_0)}^3$],
the volume of droplet can be expressed as
$V = 4\pi R_0^3 ( 1 - \Delta W/W_0 + v_c )/3$ from volume conservation.
The total Gibbs free energy of the system can be written as
\begin{align}\label{Gibbs}
  &&F_\mathrm{tot}(u_\alpha, w, v_c; \Delta W) =
  4\pi R_0^2 \gamma v_c^{2/3}+  \frac{4\pi R_0^3}{3}\Delta p_{\mathrm{lc}} v_c\\
&&\quad\quad+ \int  \mathrm{d}S  \left[\frac{1}{2} \left( E_{\alpha\beta} N^{\alpha\beta} + K_{\alpha\beta}
    M^{\alpha\beta} \right)  -  \Delta p_{\mathrm{ol}} w \vphantom{\frac{1}{2}}\right]\nonumber,
\end{align}
where $\Delta p_{\mathrm{ol}}=p_{\mathrm{o}}-p_{\mathrm{l}}$ and $\Delta
p_{\mathrm{lc}}=p_{\mathrm{l}}-p_{\mathrm{c}}$, with $p_{\mathrm{o}}\sim
10^{5}~\mathrm{Pa}$, $p_{\mathrm{l}}$ and $p_{\mathrm{c}}$ as the pressure of the outside, the liquid and the cavity, respectively.
The pressure of the cavity can be expressed as
$p_{\mathrm{c}}=-k_{B}T/v_{l} \cdot W_{0} (-\frac{p_{\mathrm{eq}}
    v_{l}}{k_{B}T}e^{-\frac{p_{\mathrm{eq}} v_{l}}{k_{B}T}-\frac{2 \gamma
      v_{l}}{R_c k_{B}T}})$~\cite{supple},
where $v_l$ is the volume of a single liquid molecule, $W_{0}(x)$ is Lambert \emph{W} function and $p_{\mathrm{eq}}$ is the vapor pressure in the bulk.
Taking typical values of $v_l\sim 10^{-29}~\mathrm{m}^{3}$ and $p_{\mathrm{eq}}\sim
10^{5}~\mathrm{Pa}$, we have $p_{\mathrm{eq}}v_l/ (k_\mathrm{B}T)\sim 10^{-3}$, and then the pressure of cavity can be approximated by $p_\mathrm{c} \approx
p_\mathrm{eq}\exp[- 2\gamma v_l/(R_c k_\mathrm{B} T)]$.
In Equation~(\ref{Gibbs}), the first term on the right-hand-side (RHS) denotes the interfacial energy of the interface between cavity and liquid, the second term denotes the work done by the pressure difference between the liquid and the cavity,
and the last term denotes the elastic energy of the gel layer.
In the last term, $E^{\alpha\beta}$ and $K^{\alpha\beta}$ are the stretching and the bending strains, respectively, $N^{\alpha\beta}$
and $M^{\alpha\beta}$ are the stretching stress and the bending moment, respectively, $w$ is the normal displacement along the radial direction of the gel layer.
Note that we have taken inward normal displacement as positive.
Here we adopt the Donnell-Mushtari-Vlasov (DMV) strain-displacement relations,
which are valid for small deformations, as
$E_{\alpha\beta} =  (\nabla_\beta u_\alpha + \nabla_\alpha u_\beta)/2 - b_{\alpha\beta} w + \partial_\alpha w\, \partial_\beta w/2$ and $K_{\alpha\beta} = \nabla_\alpha \nabla_\beta w$,
where $\nabla$ is covariant derivative, $u_\alpha$ is are the tangential displacement along $\alpha$ direction and $b_{\alpha\beta}$ is curvature tensor of the gel layer shell~\cite{niordson1985}.
The stretching stress $N^{\alpha\beta}$ and the bending moment
$M^{\alpha\beta}$ can be expressed as functions of the strain tensors:
$N^{\alpha\beta}= Eh [(1-\nu)E^{\alpha\beta} + \nu g^{\alpha\beta} E^\gamma_\gamma]/(1-\nu^2)$
and
$M^{\alpha\beta} = Eh^3[(1-\nu) K^{\alpha\beta} + \nu g^{\alpha\beta} K^\gamma_\gamma]/[12(1-\nu^2)]$,
with $g^{\alpha\beta}$ being the metric tensor of shell surface, and $h$, $E$
and $\nu$ being the thickness, Young's modulus and the Poisson's ratio of the
gel layer, respectively~\cite{niordson1985}.

\paragraph{Critical wrinkling pressure.}
The gel layer can wrinkle or buckle during the drying process,
and we first discuss the criterion of when wrinkling or buckling can occur in the droplet.
If there is no cavity, the total energy reduces to the purely elastic one [last term on RHS of Equation~(\ref{Gibbs})].
To obtain the critical wrinkling pressure,
we can perform the linear instability analysis on the Euler-Lagrange (EL) equations obtained from the optimisation of the elastic  energy~\cite{supple}.
First, the uniform solution of EL equations is $u_\phi = u_\theta = 0$,
$w_0 =(1-\nu) \Delta p_{\mathrm{ol}} R_0^2/(2Eh)$,
which represents the uniform contraction of the gel layer under the pressure
difference between outside and liquid $\Delta p_{\mathrm{ol}}$.
By adding a small perturbation $w_1$ in the normal displacement
with the total modified normal displacement as $w=w_0+w_1$ and introducing a small perturbation $\chi_1$ in the Airy stress function~\cite{niordson1985},
we can obtain
\begin{subequations}\label{perturbation}
\begin{align}
  &\frac{Eh^3}{12 {\left( 1-\nu^2 \right)}} \Delta^2 w_1 - \frac{1}{R_0} \Delta
  \,\chi_1 + \frac{\Delta p_{\mathrm{ol}} R_0}{2}\Delta w_1 = 0, \\
  & \frac{1}{Eh} \Delta^2 \, \chi_1 + \frac{1}{R_0} \Delta w_1 = 0,
\end{align}
\end{subequations}
after keeping terms up to the linear order in the EL equations.
By expressing $w_{1}$ and $\chi_1$ with spherical harmonics $Y^m_l(\phi,\theta)$ (eigenmodes of Laplace operator), as $w_1 = A\, Y^m_l$ and $\chi_1 = B\, Y^m_l$,
then from Equation~(\ref{perturbation}) we can obtain the possible eigenmodes denoted by $l$ as a function of the pressure $ \Delta p_{\mathrm{ol}}$,
\begin{equation}
  l(l+1)= \frac{\Delta p_{\mathrm{ol}}R_{0}^{3}}{4D}\left( 1\pm \sqrt{1-
      \frac{16DEh}{(\Delta p_{\mathrm{ol}})^2 R_0^4}} \right),
\end{equation}
where $D = Eh^3/[12(1-\nu^2)]$.
The minimal pressure $p^*_{\mathrm{w}}$ to have a real and positive $l(l+1)$,
\emph{i.e.},
the critical pressure leading to wrinkling, is~\cite{niordson1985}
\begin{equation}
  p^*_{\mathrm{w}} = \frac{2E}{\sqrt{3 {\left( 1-\nu^2 \right)}}} {\left( \frac{h}{R_0}
    \right)}^2,
\end{equation}
illustrating that the gel layer can easily wrinkle if the gel layer is soft (small Young's modulus) and thickness is small compared to the droplet size.
If there is a ring-like defect with a smaller Young's modulus in the gel layer,
then buckling can easily happen and the corresponding buckling pressure can be obtained by shallow shell approximation,
which gives the same value as the critical wrinkling pressure $p^*_{\mathrm{w}}$~\cite{supple}.
Note this critical wrinkling pressure is obtained for perfect sphere and is
sensitive to imperfections~\cite{thompson2015,hutchinson2016}. If
imperfections are present in shell possibly due to density
inhomogeneity~\cite{datta2012}, orthotropic elasticity~\cite{munglani2019},
\emph{etc.}, the critical wrinkling pressure can drop a lot by an empirical
proportion. Nevertheless, the above critical wrinkling pressure can still dominate the wrinkling behavior.
In the following discussions, we will keep the term `wrinkling instability' for denoting both wrinkling and buckling instability without losing generality.
\begin{figure*}[htpb]
  \centering
  \includegraphics[width=0.9278\linewidth]{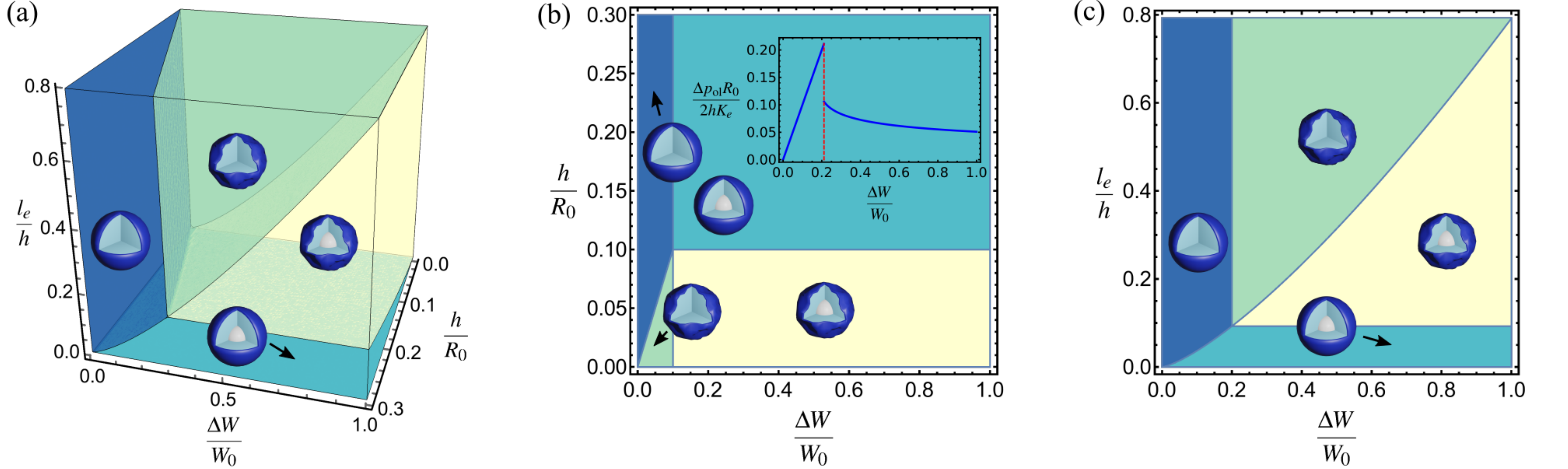}
  \caption{\label{fig:phase_diagram}(a) 3D phase diagram with three
    dimensionless varying variables, \emph{i.e.}, elastic length over shell
    thickness $l_e/h$, shell thickness over initial droplet radius $h/R_0$, and
    the reduced weight loss of the droplet $\Delta W/ W_0$. Here Poisson ratio $\nu =
    0.5$. (b) Cross section of 3D phase diagram with fixed $l_e/h=0.037$
    demonstrating one of the two drying pathways. Namely when the shell is thick
    enough, cavity develops with increase of weight loss. Inset: the pressure
    difference between outside and liquid is smaller than the critical
    cavitation pressure and constantly
    decreases after cavitation, impeding the occurrence of wrinkling.
    In the inset, $l_e/h=0.1$.
    (c) Cross section of 3D phase diagram with fixed $h/R_0=0.2$ showing the other drying
    pathway, \emph{i.e.}, from gel layer to wrinkling to both wrinkling and
    cavitation with high elastic length $l_e$.
  }
\end{figure*}

\paragraph{Critical cavitation pressure.}
Meanwhile, there is a critical cavitation pressure $p^*_\mathrm{c}$ supposing that there is no wrinkling.
In this case,
the elastic energy of the gel layer is simply $F_\mathrm{elastic} = 12\pi h K_{e} w_0^2$~\cite{meng2014},
where $K_{e}=E/[3(1-\nu)]$ is an effective elastic modulus.
After the insertion of the mass conservation relation,
the total Gibbs free energy of the system is
\begin{align}\label{eq:helmholtz}
    F_\mathrm{tot}(v_c; \Delta W) \!=& \,4\pi R_0^2 \,\gamma\, v_c^{2/3} +
    \frac{4\pi R_0^3}{3}\Delta p_{\mathrm{oc}} v_c\\\nonumber
    &+\frac{4\pi R_0^2 \,h K_{e}}{3} {\left( v_c - \frac{\Delta W}{W_0} \right)}^2,
\end{align}
where $\Delta p_{\mathrm{oc}} =  p_\mathrm{o} - p_\mathrm{c}$.
The critical point for the above total Gibbs free energy to have a non-zero local
minimum point is determined by $\partial F_\mathrm{tot} / \partial v_c =0$ and
$F_{\mathrm{tot}}(v_c) = F_{\mathrm{tot}}(0)$,
from which we can obtain the critical cavity volume as $v_c^* =
[l_e/(2h)]^{3/4}$ with the elastic length defined by $l_e \equiv 2\gamma / K_e$
and critical weight loss for cavitation as $\Delta W^*/W_0 = [2 +
p_{\mathrm{eq}}v_l/(k_\mathrm{B} T)] v_c^*$~\cite{meng2014}.
Utilisation of the mass conservation under uniform contraction with no cavity
gives us the relation between pressure and weight loss
$\Delta p_{\mathrm{ol}} = 2hK_{e}/R_0 \cdot\Delta W/W_0.$
Then the critical cavitation pressure is
\begin{equation}
  \label{eq:critial_cav_pressure}
  p^*_\mathrm{c} = \frac{4hK_e}{R_0}\left( 1 + \frac{p_{\mathrm{eq}}v_l}{2 k_\mathrm{B} T}\right) {\left( \frac{l_e}{2h} \right)}^{3/4}.
\end{equation}

By comparing the critical wrinkling pressure $p^*_{\mathrm{w}}$ and the
critical cavitation pressure $p^*_{\mathrm{c}}$,
we can obtain the criterion for wrinkling happening ahead of the droplet cavitation at the early stage of the drying process, as
\begin{align}
  \left( \frac{2}{9}\right)^{1/4}\!\! \left(\frac{1+\nu}{1-\nu}\right)^{1/2}
  \!\!\left(
    1 + \frac{p_{\mathrm{eq}}v_l}{2 k_\mathrm{B} T}\right) \! \frac{R_0}{h} \left(
    \frac{l_e}{h}\right)^{3/4} \! > 1;
\end{align}
otherwise, cavitation occurs first.
In other words, if the gel layer is thin and soft, the droplet tends to wrinkle at the surface.
More importantly,
such divergence in the onset instabilities at the early stage of the drying process leads to different drying pathways and can determine the final morphology of the drying droplet as discussed below.

\paragraph{Drying pathway--no wrinkling after cavitation.}
First we investigate whether wrinkling still occurs if cavitation happens first.
This can be answered by understanding how the pressure difference between
outside and liquid $\Delta p_{\mathrm{ol}}$ changes with the increase of weight loss $\Delta W$.
As shown in the inset of Fig.~\ref{fig:phase_diagram} (b), $\Delta p_{\mathrm{ol}}$
constantly increases before cavitation and becomes $
2\gamma[1 + p_{\mathrm{eq}}v_l/(k_\mathrm{B} T)]v_c^{-1/3}/R_0$ after cavitation.
At the critical cavitation point, $\Delta p_{\mathrm{ol}} = 2hK_e/R_0\cdot
[1+p_{\mathrm{eq}} v_l/(k_\mathrm{B}T)][l_e/(2h)]^{3/4}$, which is smaller than the critical
cavitation pressure $p_\mathrm{c}^*$.
Meanwhile, the first derivative of $\Delta p_{\mathrm{ol}}$ after cavitation with regards to the weight loss,
$\partial \Delta p_{\mathrm{ol}}/\partial \Delta W = - 2\gamma / (R_0W_0)\cdot
1/[3[1+p_{\mathrm{eq}} v_l/(k_\mathrm{B}T)]^{-1}v_c^{4/3} - v_c^{*4/3}] < 0$,
means that $\Delta p_{\mathrm{ol}}$ decreases further with increasing weight loss $\Delta W$,
indicating no wrinkling can happen after cavitation since the critical wrinkling pressure can not be reached any more [drying pathway in Fig.~\ref{fig:schematics_phases} as (a)$\rightarrow$ (b) $\rightarrow$ (c1) $\rightarrow$ (c2)].

\paragraph{Drying pathway--cavitation after wrinkling.}
We proceed to investigate whether the cavity can still form if wrinkling occurs first.
By assuming wrinkling as a perturbation of a single spherical harmonic mode
and the perturbation amplitude is small,
then the Helmholtz free energy can be expressed in the same form as that in
equation~(\ref{eq:helmholtz}),
but with a modified effective elastic modulus~\cite{supple}
\begin{align}
  K'_e = K_e {\left[1- \frac{3(1+\nu) A'^2}{8\pi} \right]},
\end{align}
where $A'=A/h$ is the ratio of the perturbation amplitude over the gel layer thickness.
Then the problem of whether cavitation can happen after wrinkling, is reduced
to if cavitation can happen under uniform contraction with the new elastic modulus
$K'_e$.
Then we can obtain the new critical weight loss for cavitation,
namely, $\Delta W'/W_0 = [ l'_e/(2h)] ^{3/4}$,
where $l'_e \equiv 2\gamma/K'_e$ is the new elastic length.
Thus cavitation can still happen after wrinkling,
when the weight loss exceeds the above critical value.
Meanwhile, the effective elastic modulus $K'_e$ is smaller than the
one without wrinkling $K_e$,
resulting in a larger value of the critical weight loss than that in the case without wrinkling.
Note that according to the previous discussion,
the pressure difference between outside and liquid $\Delta p_{\mathrm{ol}}$ drops at cavitation and
constantly decreases after cavitation with
the increase of weight loss [see inset of Fig.~\ref{fig:phase_diagram}(b)],
thus the wrinkles may disappear due to the decrease of $\Delta p_{\mathrm{ol}}$.
However, in practical drying process, the formed wrinkles of the gel layer can become rigid, \emph{e.g.} turning glassy~\cite{pauchard2003},
and the decrease of $\Delta p_{\mathrm{ol}}$ induced by the cavitation is not
high enough to flatten the wrinkles on the gel layer with increased rigidity.
In this case,
the wrinkles can still remain regardless of the decrease of pressure difference
between outside and liquid after cavitation [drying pathway in Fig.~\ref{fig:schematics_phases} as (a)$\rightarrow$ (b) $\rightarrow$ (d1) $\rightarrow$ (d2)].
Note that the gel layer can still be permeable to solvent despite of
rigidification. Thus the subsequent evaporation and cavity enlargement are not hindered.

In Fig.~\ref{fig:phase_diagram}(a),
a 3D phase diagram summarising the above discussions of the drying pathways of a soft matter droplet is provided in the space of (elastic length, gel layer thickness and weight loss),
together with its two cross sections of given fixed elastic length in Fig.~\ref{fig:phase_diagram}(b) and of given fixed thickness in Fig.~\ref{fig:phase_diagram}(c), respectively.
Note that the weight loss in the phase diagram can play the role of time if the
evaporation rate of the droplet is known~\cite{boraey2014,meng2016}.
At the early stage of the drying process, \emph{i.e.}, when $\Delta W$ is small,
either wrinkling or cavitation instability can occur,
resulting from the competition between the surface energy of the cavity,
and the bending and the stretching elastic energy of the gel layer.
When the gel layer is thin and soft [small $h$ and small $K_{e}$ (large $l_e$)],
then wrinkling occurs, which is obvious in Fig.~\ref{fig:phase_diagram}(b) and Fig.~\ref{fig:phase_diagram}(c); otherwise, cavitation happens.
Meanwhile,
the choice in either wrinkling or cavitation instability at the early stage of the drying process,
determines the later morphology evolution and the final product of the drying soft matter droplet: (a) if wrinkling takes place ahead of the cavitation, then there will still be cavity formed in the droplet with the ongoing evaporation of the solvents and the final configuration of the drying droplet will be a hollow and wrinkled particle;
(b) if cavitation happens first, then there will be no wrinkling in the later
drying process due to the decreasing pressure difference between the outside and the droplet,
and a spherical shell is left after finishing the whole evaporation process.

We here investigate how the morphology of a drying soft matter droplet can evolve with time based on a pseudo-dynamic analysis.
The elastic properties of the gel layer formed at the surface of the soft matter droplet play the key role in determining the drying pathways of the droplet, including both the instabilities triggered at the early stage of the drying process, the later morphology evolution, and the final configurations.
Although simplifications like the non-evolving gel layer have been made for analytic discussions,
we believe this portable model has captured the essential physics underlying the morphology evolution of a drying soft matter droplet, which can guide the industrial fabrications of micro-particles with desired morphologies and functions.

\begin{acknowledgments}
F.Y. acknowledges the support of the National Natural Science Foundation of China (Grant No. 11774394) and the Key Research Program of Frontier Sciences of Chinese Academy of Sciences (Grant No. QYZDB-SSW-SYS003). G.D. thanks Xiao Lin for fruitful discussions.
\end{acknowledgments}

\end{document}



\title{Drying Pathways of an Evaporating Soft Matter Droplet\\ \normalfont\textit{Supplemental Material}}

\author{Guangle Du}
\affiliation{CAS Key Laboratory of Theoretical Physics, Institute of Theoretical
Physics,Chinese Academy of Sciences, Beijing 100190, China}
\affiliation{Wenzhou Institute, University of Chinese Academy of Sciences, Wenzhou, Zhejiang 325001, China}
\affiliation{School of Physical Sciences, University of Chinese Academy of Sciences, Beijing 100049, China}

\author{Fangfu Ye}
\affiliation{Wenzhou Institute, University of Chinese Academy of Sciences, Wenzhou, Zhejiang 325001, China}
\affiliation{School of Physical Sciences, University of Chinese Academy of Sciences, Beijing 100049, China}
\affiliation{Beijing National Laboratory for Condensed Matter Physics, Institute of Physics, Chinese Academy of Sciences, Beijing 100190, China}
\affiliation{Songshan Lake Materials Laboratory, Dongguan, Guangdong 523808, China}

\author{Masao Doi}
\affiliation{Center of Soft Matter Physics and its Applications, Beihang University, Beijing 100191, China}

\author{Fanlong Meng}
\email{fanlong.meng@itp.ac.cn}
\affiliation{CAS Key Laboratory of Theoretical Physics, Institute of Theoretical
Physics,Chinese Academy of Sciences, Beijing 100190, China}

\maketitle


\section{Derivation of Euler-Lagrange equations}\label{sec:el}
In this section, we give the detailed derivation of Euler-Lagrange equations
from the elastic free energy of shell
\begin{equation}
  F_\text{\text{elastic}}(u_\alpha, w) = \int_S \mathrm{d}S \left[ \frac{1}{2} {\left(
        E_{\alpha\beta} N^{\alpha\beta} + K_{\alpha\beta} M^{\alpha\beta}
      \right) - \Delta p_{\mathrm{ol}} w}\right],
  \label{eq:elastic_energy}
\end{equation}
where $\Delta p_{\mathrm{ol}}$ is pressure difference between outside and
liquid, \emph{i.e.}, net pressure exerted on the shell.
$u_\alpha$ and $w$ are, respectively, tangential and normal
displacements on the gel layer shell.
$E_{\alpha\beta}$ and $K_{\alpha\beta}$ are,
respectively, the stretching and bending strains. $N^{\alpha\beta}$ and
$M^{\alpha\beta}$ are stretching stress and bending moment, which can be written
in terms of stretching and bending strains
\cite{niordson1985}
\begin{align}
  N^{\alpha\beta} &= \frac{Eh}{1-\nu^2} {\left[ {\left( 1-\nu \right)}
      E^{\alpha\beta} + \nu g^{\alpha\beta} E^\gamma_\gamma \right]},\\
  M^{\alpha\beta} &= \frac{Eh^3}{12 {\left( 1-\nu^2 \right)}} {\left[ {\left(
          1-\nu \right)} K^{\alpha\beta} + \nu g^{\alpha\beta} K^\gamma_\gamma
    \right]},
\end{align}
with $g^{\alpha\beta}$ being metric tensor of shell surface and $h$, $E$ and
$\nu$ being thickness, Young's modulus and Poisson's ratio of shell.
We adopt here the Donnell–Mushtari–Vlasov (DMV) strain-displacement relations
\cite{niordson1985}
\begin{align}
  \label{eq:stretching}
  E_{\alpha\beta} &= \frac{1}{2} {\left( \nabla_\beta u_\alpha + \nabla_\alpha
      u_\beta\right)} - b_{\alpha\beta} w + \frac{1}{2} \partial_\alpha w\,
  \partial_\beta w, \\
  \label{eq:bending}
  K_{\alpha\beta} &= \nabla_\alpha \nabla_\beta w,
\end{align}
where $b_{\alpha\beta}$ is curvature tensor and $\nabla$ is covariant derivative.
Note in the elastic free energy Eq.~(\ref{eq:elastic_energy}), there is second order
derivative of $w$. In general curved space, the Euler-Lagrange equations are
\begin{align}
  & \frac{\partial f_\text{elastic}}{\partial u_\alpha} - \nabla_\beta
  \frac{\partial f_\text{elastic}}{\partial \nabla_\beta u_\alpha} = 0, \\
  & \frac{\partial f_\text{elastic}}{\partial w} - \nabla_\alpha \frac{\partial
    f_\text{elastic}}{\partial \partial_\alpha w} + \nabla_\beta \nabla_\alpha
  \frac{\partial f_\text{elastic}}{\partial \partial_\alpha\partial_\beta w} =
  0,
\end{align}
where $f_\text{elastic}$ is the elastic free energy density. Then
\begin{align}
  & \nabla_\beta N^{\alpha\beta} = 0, \\
  & - N^{\alpha\beta} b_{\alpha\beta} - \Delta p_{\mathrm{ol}} - N^{\alpha\beta} \nabla_\alpha
  \nabla_\beta w + \nabla_\beta \nabla_\alpha M^{\alpha\beta} =
  0.\label{eq:el_w}
\end{align}
An auxiliary function called Airy stress function can be introduced to reduce
the above three equations into two equations. Let
$
  N^{\alpha\beta} \equiv \varepsilon^{\alpha\gamma}\varepsilon^{\beta\delta}
  \nabla_\gamma \nabla_\delta \, \chi.
$
According to the formula
\begin{align}
  \nabla_\alpha \nabla_\beta A^\gamma - \nabla_\beta \nabla_\alpha A^\gamma =
  A^\delta R^\gamma_{\;\;\delta\alpha\beta},
\end{align}
where $R^{\alpha}_{\;\;\beta\gamma\delta}$ is Riemann curvature tensor and on 2D surface
\begin{align}
  R^{\alpha}_{\;\;\beta\gamma\delta} = K {\left( g^\alpha_\gamma g_{\beta\delta} -
      g^\alpha_\delta g_{\beta\gamma} \right)},
\end{align}
where $K$ is Gauss curvature,
\begin{align}
  \nabla_\alpha\nabla_\beta A^\gamma = \frac{1}{2} {\left( \nabla_\alpha
      \nabla_\beta + \nabla_\beta \nabla_\alpha \right)} A^\gamma + \frac{1}{2}
  K {\left(A_\beta g^\gamma_\alpha - A_\alpha g^\gamma_\beta\right)}.
\end{align}
Denote the characteristic length of deformation with $l$. The first symmetric part on the right hand side is of order $1/l^2$ and the
second anti-symmetric part is of order $K$. If we assume $1/l^2 \gg K$,
\emph{i.e.}, the inverse of characteristic length of deformation squared is far
larger than Gauss curvature, then the second anti-symmetric part can be omitted
and only the first symmetric part remains. Under this assumption, covariant
derivatives commute, $\nabla_\alpha \nabla_\beta = \nabla_\beta \nabla_\alpha$. Hence
\begin{align}
  \nabla_\alpha N^{\alpha\beta} &= \varepsilon^{\alpha\gamma} \nabla_\alpha
  \nabla_\gamma {\left( \varepsilon^{\beta\delta} \nabla_\delta \,\chi \right)}
  = 0.
\end{align}
Consequently, Eq.~(\ref{eq:el_w}) becomes
\begin{equation}
   \frac{Eh^3}{12 {\left( 1-\nu^2 \right)}} \Delta^2 w - b_{\alpha\beta}
  \varepsilon^{\alpha\gamma} \varepsilon^{\beta\delta}
  \nabla_\gamma\nabla_\delta\, \chi -
  \varepsilon^{\alpha\gamma}\varepsilon^{\beta\delta} \nabla_\alpha
  \nabla_\beta \, w\nabla_\gamma \nabla_\delta\, \chi - \Delta p_{\mathrm{ol}} = 0. \label{eq:el1}
\end{equation}
The Airy stress
function fulfills an additional compatibility equation.
The stretching strain can be written in terms of stretching stress
\begin{align}
  E_{\alpha\beta} = \frac{1}{Eh} {\left[ {\left( 1+\nu \right)} N_{\alpha\beta}
      - \nu g_{\alpha\beta} N^\gamma_\gamma\right]}. \label{eq:E1}
\end{align}
Acting $\varepsilon^{\alpha\gamma} \varepsilon^{\beta\delta} \nabla_\gamma
\nabla_\delta$ on Eqs.~(\ref{eq:E1}) and (\ref{eq:stretching}), we obtain the compatibility equation
\begin{equation}
  \frac{1}{Eh} \Delta^2 \chi + b_{\alpha\beta}
  \varepsilon^{\alpha\gamma} \varepsilon^{\beta\delta}
  \nabla_\gamma\nabla_\delta\, w + \frac{1}{2}\,
  \varepsilon^{\alpha\gamma}\varepsilon^{\beta\delta} \nabla_\alpha
  \nabla_\beta \, w\nabla_\gamma \nabla_\delta\, w = 0. \label{eq:el2}
\end{equation}
For spherical shell, $b_{\alpha\beta} = g_{\alpha\beta}/R_0$. The
Euler-Lagrange Eqs.~(\ref{eq:el1}) and (\ref{eq:el2}) become~\cite{niordson1985}
\begin{align}
  & \frac{Eh^3}{12 {\left( 1-\nu^2 \right)}} \Delta^2 w - \frac{1}{R_0}\Delta\, \chi -
  \varepsilon^{\alpha\gamma}\varepsilon^{\beta\delta} \nabla_\alpha
  \nabla_\beta \, w\nabla_\gamma \nabla_\delta\, \chi - \Delta p_{\mathrm{ol}} = 0,
  \label{eq:el1_sph}\\
  & \frac{1}{Eh} \Delta^2 \chi + \frac{1}{R_0}\Delta\, w + \frac{1}{2}\,
  \varepsilon^{\alpha\gamma}\varepsilon^{\beta\delta} \nabla_\alpha
  \nabla_\beta \, w\nabla_\gamma \nabla_\delta\, w = 0, \label{eq:el2_sph}
\end{align}
which completes the derivation of the Euler-Lagrange equations.

\section{Linear stability analysis of Euler-Lagrange equations}\label{sec:uniform}
In this section, we perform linear stability analysis on
Eqs.~(\ref{eq:el1_sph}) and (\ref{eq:el2_sph}) to find the critical wrinkling
pressure.
For uniform solution, Eq.~(\ref{eq:el1_sph}) reduces to
\begin{align}
  - \Delta\, \chi_0 = \Delta p_{\mathrm{ol}}R_0.
\end{align}
Recall that
$
  N^{\alpha\beta} \equiv \varepsilon^{\alpha\gamma} \varepsilon^{\beta\delta}
  \nabla_\gamma \nabla_\delta \,\chi.
$
Then
\begin{align}
  N^\alpha_{0\alpha} = \varepsilon^{\alpha\gamma} \varepsilon_{\alpha\delta}
  \nabla_\gamma \nabla^\delta \,\chi_0 = \Delta \,\chi_0.
\end{align}
There should be $N^\phi_{0\phi} = N^\theta_{0\theta}$ for spherical shell in uniform
solution. So
\begin{align}
  N^\phi_{0\phi} = N^\theta_{0\theta} = - \frac{1}{2} \Delta p_{\mathrm{ol}} R_0.
\end{align}
According to stretching strain-stress relation Eq.~(\ref{eq:E1}) and the DMV strain-displacement
relation Eq.~(\ref{eq:stretching}),
\begin{align}
  E_{\theta\theta} &= \frac{1}{Eh} {\left[ {\left( 1+\nu \right)}
      N_{\theta\theta} - \nu g_{\theta\theta} N^\gamma_\gamma \right]}
  = -\frac{1}{R_0} g_{\theta\theta} w_0.
\end{align}
Then
\begin{align}
  w_0 &= - \frac{R_0}{Eh} {\left[ {\left( 1+\nu \right)} N^\theta_\theta - \nu
      N^\gamma_\gamma\right]}
  = \frac{ {\left( 1-\nu \right)} \Delta p_{\mathrm{ol}} R_0^2}{2Eh}.
\end{align}

\begin{figure}[tb]
  \centering
  \includegraphics[width=0.6\linewidth]{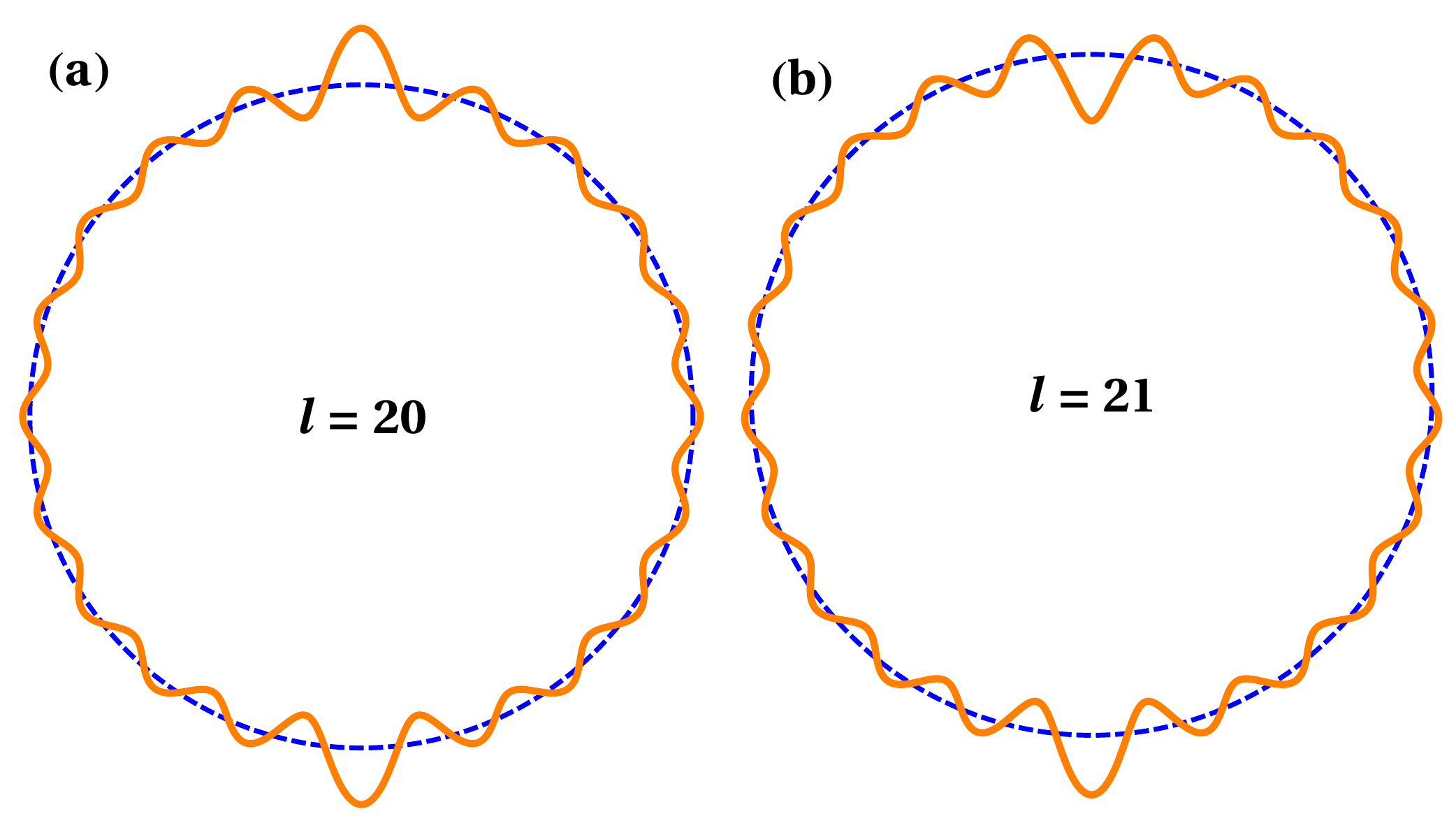}
  \caption{\label{fig:wrinkling_modes} Schematics of typical wrinkling modes with $\nu=0.5$.
    (a) Symmetric mode with $l=20$ and $R_0/h = 140$. (b) Anti-symmetric mode
    with $l=21$ and $R_0/h= 154$.}
\end{figure}

To find the critical pressure leading to wrinkling, we
perform linear stability analysis on Eqs.~(\ref{eq:el1_sph}) and (\ref{eq:el2_sph}).
Expanding $w = w_0 + w_1$
and $\chi = \chi_0 + \chi_1$ and keeping terms up to linear order, we obtain
\begin{align}
  &\frac{Eh^3}{12 {\left( 1-\nu^2 \right)}} \Delta^2 w_1 - \frac{1}{R_0} \Delta
  \,\chi_1 + \frac{\Delta p_{\mathrm{ol}} R_0}{2}\Delta w_1 = 0, \\
  & \frac{1}{Eh} \Delta^2 \, \chi_1 + \frac{1}{R_0} \Delta w_1 = 0.
\end{align}
Note spherical harmonics $Y^m_l(\phi,\theta)$ are eigenmodes of Laplace
operator $\Delta$. Substitution of $w_1 = A\, Y^m_l$ and $\chi_1 = B\, Y^m_l$
into the above equations gives us
\begin{align}
  & {\left[ \frac{Eh^3}{12\left(1-\nu^2\right)} x - \frac{\Delta p_{\mathrm{ol}}R_0}{2} \right]} A +
  \frac{1}{R_0} B = 0, \\
  & - \frac{1}{R_0} A + \frac{1}{Eh} x B = 0,
\end{align}
where $x \equiv l(l+1)/R_0^2$.
To have non-zero solution of $A$ and $B$, there must be
\begin{align}
  \frac{Eh^3}{12 {\left( 1-\nu^2 \right)}} x^2 - \frac{\Delta p_{\mathrm{ol}}R_0}{2} x +
  \frac{Eh}{R_0^2} = 0.
\end{align}
The minimal $\Delta p_{\mathrm{ol}}$ to have a real and positive $x$
corresponds to the critical wrinkling pressure~\cite{niordson1985}
\begin{align}
  p^*_{\mathrm{w}} = \frac{2E}{\sqrt{3 {\left( 1-\nu^2 \right)}}} {\left( \frac{h}{R_0}
    \right)}^2.
\end{align}

In Fig.~\ref{fig:wrinkling_modes}, we have shown two typical wrinkling modes at
the critical wrinkling pressure.

\section{Derivation of the pressure in the cavity $p_{\mathrm{c}}$}
In this section, the pressure in cavity is derived.
The equilibrium condition between liquid and cavity is
\begin{align}
  \mu_c (T, p_{\mathrm{c}})  = \mu_l(T, p_{\mathrm{l}}), \quad
  p_{\mathrm{c}} = p_{\mathrm{l}} + \frac{2\gamma}{R_c},
\end{align}
where $\mu_c$, $\mu_l$, $p_{\mathrm{c}}$ and $p_{\mathrm{l}}$ are,
respectively, chemical potentials and pressures of cavity and liquid.
By assuming the gas in cavity is ideal and liquid is incompressible, we have
\begin{align}
  \mu_c(T, p_{\mathrm{c}})  =& \mu^0_c(T) + k_BT \ln p_{\mathrm{c}}, \\
  \mu_l(T, p_{\mathrm{l}}) =& \mu^0_l(T) + p_\mathrm{l} v_l,
\end{align}
where $v_l$ is the specific volume of liquid molecule.
Then
\begin{equation} \label{eq:equilibrium}
  \mu_c^0(T) + k_B T \ln p_\mathrm{c} = \mu^0_l(T) + \left(p_\mathrm{c} -
    \frac{2\gamma}{R_c}\right) v_l.
\end{equation}
Let $p_{\mathrm{eq}}(T)$ be the solution of Eq.~(\ref{eq:equilibrium}) when $R_c
\to \infty$. Eq.~(\ref{eq:equilibrium}) becomes
\begin{equation}
  k_BT \ln \left( \frac{p_\mathrm{c}}{p_{\mathrm{eq}}}\right) =
  (p_{\mathrm{c}} - p_{\mathrm{eq}})v_l  - \frac{2\gamma}{R_c} v_l.
\end{equation}
Thus
\begin{equation} \label{eq:pc}
  p_{\mathrm{c}} = - \frac{k_BT}{v_l} W_0 \left(- \frac{
      p_{\mathrm{eq}}v_l}{k_BT} \mathrm{e}^{ - \frac{ p_{\mathrm{eq}}v_l}{k_BT} -
      \frac{2\gamma v_l}{R_c k_B T}}\right),
\end{equation}
where $W_0$ is Lambert \emph{W} function.
Note that the specific volume of water molecule is $3\times
10^{-29}\mathrm{m}^3$. So $p_{\mathrm{eq}}v_l / (k_B T) \sim 10^{-3}$ and
Eq.~(\ref{eq:pc}) can be approximated by
\begin{equation}
  p_{\mathrm{c}} \approx p_{\mathrm{eq}} \exp\left( -\frac{2\gamma v_l}{R_c k_B T}\right).
\end{equation}

We now estimate the order of the exponential argument.
Recall that the surface tension of distilled water at room temperature is $\gamma = 7.2 \times 10^{-2} \mathrm{N}/\mathrm{m}$.
The specific volume of water molecule is $v_{l} = 3 \times 10^{-29} \mathrm{m}^{3}$.
The effective osmotic modulus is of the order $K_{e} \sim 1$ GPa.
The typical initial radius of droplet in experiment $R_{0}\sim 10^{-6}$m.
If we assume the thickness of shell is $h=0.01R_{0}$,
then $2\gamma v_{l} /(R_{c}k_{B}T) \sim 10^{-2.75} \ll 1$.
If we allow $K_{e}$ and $h$ to vary and require $2\gamma v_{l}/(R_{c}k_{B}T)  < 0.1$,
then $v_{c} > 10^{-6}$ or equivalently $l_{e}/(2h) > 10^{{-8}}$, which should be easily fulfilled.
Hence
\begin{equation}
\label{eq:7}
p_{\mathrm{c}} \approx p_{\text{eq}} \left(1-\frac{2\gamma v_{l}}{R_{c} k_{B}T}\right).
\end{equation}
Assume $p_{\mathrm{o}}=p_{\text{eq}} = 1$ atm. Then $\Delta p_{\mathrm{oc}} = 2\gamma p_{\text{eq}} v_{l}/(R_{c} k_{B}T)$.

\section{Derivation of elastic free energy under uniform contraction}
In this section, we derive the elastic free energy under uniform contraction
before wrinkling and cavitation.
Under uniform contraction with tangential displacement zero and normal
displacement $w_0$, the bending strain is zero and stretching strain reduces to
$E_{\alpha\beta} = -g_{\alpha\beta} w_0/R_0 $, according to the DMV
strain-displacement relations in Eqs.~(\ref{eq:stretching}) and (\ref{eq:bending}).
Then Helmholtz elastic free energy is
\begin{align}
F_\mathrm{elastic} = \int_S \mathrm{d} S \frac{Eh}{2(1-\nu^2)}
\left[(1-\nu)E^{\alpha\beta}E_{\alpha\beta} + \nu
  \left(E^\gamma_\gamma\right)^2\right] = 12\pi h K_e w_0^2,
\end{align}
where an effective osmotic modulus $K_e = E/[3(1-\nu)]$ is introduced.

\section{Derivation of Helmholtz free energy with wrinkling}
\label{sec:helmholts_buckling}
In this section, we give the detailed derivation of Helmholtz free energy
after wrinkling. The wrinkling is assumed to be in the eigenmode
$w = w_0 + A Y^m_l$ and the variation of tangential displacement $\nabla_\alpha
u_\beta$ is small. Substitution of this solution into
the total Helmholtz free energy of the system leads to
\begin{align}
  \label{eq:helmholtz_buckling_raw}
  F_\text{tot} = 12\pi K_e h w_0^2 + A^2 \left[
    \frac{h^3 K_e l^2(l+1)^2}{8(1+\nu)R_0^2} - \frac{3 h K_e w_0l(l+1)}{2 R_0}
  \right]
    + 4\pi R_0^2 \gamma v_c^{2/3}.
\end{align}
In the linear instability analysis, we know
\begin{align}
  \frac{l(l+1)}{R_0^2} = \frac{\Delta p_{\mathrm{ol}}R_0}{4D} {\left( 1\pm \sqrt{1-
        \frac{16DEh}{(\Delta p_{\mathrm{ol}})^2R_0^4}} \right)},
\end{align}
where $D \equiv Eh^3/[12(1-\nu^2)]$. The pressure of near wrinkling onset is
$\Delta p_{\mathrm{ol}}\simeq p_{\mathrm{c}}$. Then expansion around $\Delta
p_{\mathrm{ol}} = p^*_{\mathrm{c}}$ and keep up to linear term of
$\Delta p_{\mathrm{ol}}-p_{\mathrm{c}}^*$ gives us
\begin{align}
  \frac{l(l+1)}{R_0^2} \simeq \frac{\Delta p_{\mathrm{ol}} R_0}{4D}.
\end{align}
The mass conservation relation reads
\begin{align}
  \int_S w\, \mathrm{d}S = 4\pi R_0^2 w_0 = \frac{4\pi R_0^3}{3} {\left(
      \frac{\Delta W}{W_0} - v_c \right)}.
\end{align}
Recall that $w_0 = \left.(1-\nu) \Delta p_{\mathrm{ol}} R_0^2 \middle/ (2Eh)\right.$. Then
\begin{align}
  \frac{l(l+1)}{R_0^2} = \frac{2(1+\nu)}{h^2} {\left( \frac{\Delta W}{W_0} -
      v_c\right)}.
\end{align}
Finally we arrive at
\begin{align}
  F_\text{tot} = \frac{4\pi R_0^2 h K'_e}{3} {\left( v_c - \frac{\Delta W}{W_0}
    \right)}^2 + 4\pi R_0^2 \gamma v_c^{2/3},
  \label{eq:helmholtz_buckling}
\end{align}
where $K'_e = K_e [1- 3(1+\nu) A'^2/(8\pi)]$ and $A' = A/h$.

\section{Occurrence conditions of phases}
Occurrence conditions of four possible phases on the drying pathways are listed as follows.
\begin{itemize}
  \item No cavitation or wrinkling:
    \begin{align}
      \frac{\Delta W}{W_0} < \min\left\{2^{1/4}\left( 1 +
          \frac{p_\mathrm{eq}v_l}{2k_BT}\right) {\left( \frac{l_e}{h}
          \right)}^{3/4},
        \;
      \sqrt{\frac{3(1-\nu)}{1+\nu}} \frac{h}{R_0}
    \right\}.
  \end{align}
  \item Cavitation only:
    \begin{align}
      & \frac{\Delta W}{W_0} > 2^{1/4}\left( 1 + \frac{p_\mathrm{eq}v_l}{2k_BT}\right) {\left( \frac{l_e}{h} \right)}^{3/4} \nonumber\\
      \text{and}\quad &
 2^{1/4}\left( 1 + \frac{p_\mathrm{eq}v_l}{2k_BT}\right) {\left( \frac{l_e}{h} \right)}^{3/4}
<\sqrt{ \frac{3(1-\nu)}{1+\nu}} \frac{h}{R_0}.
    \end{align}
  \item Wrinkling only:
    \begin{align}
 &\sqrt{\frac{3(1-\nu)}{1+\nu}} \frac{h}{R_0} < \frac{\Delta W}{W_0} <
 2^{1/4}\left( 1 + \frac{p_\mathrm{eq}v_l}{2k_BT}\right) {\left( \frac{l'_e}{h}
   \right)}^{3/4} \nonumber\\
      \text{and}\quad &
 2^{1/4}\left( 1 + \frac{p_\mathrm{eq}v_l}{2k_BT}\right) {\left( \frac{l_e}{h} \right)}^{3/4}
 >\sqrt{ \frac{3(1-\nu)}{1+\nu}} \frac{h}{R_0}.
    \end{align}
  \item Wrinkling and cavitation:
    \begin{align}
     &\frac{\Delta W}{W_0} > 
 2^{1/4}\left( 1 + \frac{p_\mathrm{eq}v_l}{2k_BT}\right) {\left( \frac{l'_e}{h}
   \right)}^{3/4} \nonumber\\
 \text{and} \quad &
 2^{1/4}\left( 1 + \frac{p_\mathrm{eq}v_l}{2k_BT}\right) {\left( \frac{l_e}{h} \right)}^{3/4}
 >\sqrt{ \frac{3(1-\nu)}{1+\nu}} \frac{h}{R_0}.
    \end{align}
\end{itemize}


%